\documentclass[11pt]{article}
\usepackage{a4wide}
\usepackage{setspace}
\usepackage{pdflscape}
\usepackage{amsmath} 
\usepackage{amssymb} 
\usepackage{subfigure} 
\usepackage{graphicx}
\usepackage{framed}
\graphicspath{{figures/}}
\usepackage{float}
\usepackage[small]{caption}
\usepackage{rotating}
\usepackage{booktabs}
\usepackage{url}
\usepackage{enumerate}
\usepackage{tabto}
\usepackage{authblk}
\usepackage{color}
\usepackage[usenames,dvipsnames,svgnames,table]{xcolor}
\usepackage{bm}
\usepackage{multirow}

\usepackage{natbib}
\usepackage{slashbox}

\def\tr{\text{tr}}

\title{A Distance-Based Test of Association Between Paired Heterogeneous Genomic Data}

\author[1,2]{Christopher Minas}
\author[3]{Edward Curry}
\author[1]{Giovanni Montana\footnote{Corresponding author: {\tt g.montana@imperial.ac.uk}}}
\affil[1]{Statistics Section, Department of Mathematics, South Kensington Campus, Imperial College London, London, SW7 2AZ}
\affil[2]{Institute of Clinical Sciences, Hammersmith Campus, Imperial College London, London, W12 0NN}
\affil[3]{Ovarian Cancer Action Research Centre, Hammersmith Campus, Imperial College London, London, W12 0NN}

\begin{document}
        
\maketitle

\begin{abstract}
Due to rapid technological advances, a wide range of different measurements can be obtained from a given biological sample including single nucleotide polymorphisms, copy number variation, gene expression levels, DNA methylation and proteomic profiles. Each of these distinct measurements provides the means to characterize a certain aspect of biological diversity, and a fundamental problem of broad interest concerns the discovery of shared patterns of variation across different data types. Such data types are heterogeneous in the sense that they represent measurements taken at very different scales or described by very different data structures. We propose a distance-based statistical test, the generalized RV (GRV) test, to assess whether there is a common and non-random pattern of variability between paired biological measurements obtained from the same random sample. The measurements enter the test through distance measures which can be chosen to capture particular aspects of the data. An approximate null distribution is proposed to compute p-values in closed-form and without the need to perform costly Monte Carlo permutation procedures. Compared to the classical Mantel test for association between distance matrices, the GRV test has been found to be more powerful in a number of simulation settings. We also report on an application of the GRV test to detect biological pathways in which genetic variability is associated to variation in gene expression levels in ovarian cancer samples, and present results obtained from two independent cohorts.      
\end{abstract}

\section{Introduction}

A proliferation in the development and application of high-throughput measurement platforms in biological research has resulted in the increasing availability of multiple levels of molecular data available for the same biological sampling units. For a range of human tumour types, for instance, the Cancer Genome Atlas (TCGA) consortium have made available genome-wide SNP genotypes, DNA copy number, genome-wide gene (and mRNA) expression, DNA methylation and proteomic profiles, as well as a range of mutation calls arising from deep sequencing. Such comprehensive molecular profiling of biological sampling unit provides opportunities for gaining insight into the mechanisms by which different molecular entities interact with one another to influence the overall state of the cell(s) in question. 
This is possible through comparing the different types of observations on each biological sampling unit, via some means of paired analysis of the individual datasets. A certain degree of complexity is inherent in this type of analysis since the measurements that are being compared are heterogeneous.

One well-studied example of the task of analysing paired heterogeneous biological data is gene expression quantitative trait loci (eQTL) mapping, which seeks to find associations between single nucleotide polymorphisms (SNPs) and transcript expression levels 
\citep{cookson2009mapping}. 
A particularly interesting application of multivariate eQTL mapping relates to the discovery of association between biomarkers for ovarian cancer, which is treated as a case study in this work.

The application of paired analysis of heterogeneous biological data types is not limited to eQTL mapping, and involves the very common problem of discovering shared patterns of variation across different biological measurements. Other areas that have seen considerable interest include the association of genetic copy number variation with gene expression levels \citep{beroukim07} 
and identification of genes for which DNA methylation is associated to gene expression \citep{louhimo2011}. 
In these and related studies, the problem of comparing heterogeneous data types can be approached using a notion of distance between pairs of biological samples. Distances are scalar-valued measures of dissimilarity between two input observations, with greater values indicating greater dissimilarity between them. For a given application, a suitable semi-metric or metric distance measure can generally be defined depending on the nature of the data and on the specific objectives of the study. Such distances computed between all pairwise combinations of samples are arranged in square, symmetric matrices. These matrices then quantify the variability in the random sample according to that particular distance measure, and can be directly compared using an appropriate statistical test. 

Distance-based approaches offer a number of advantages compared to direct comparisons involving the original measurements. Firstly, distances can overcome limitations of traditional multivariate approaches when dealing with high-dimensional random vectors observed on a much smaller number of sampling units \citep{shannon2002mantel}. Secondly, distance measures can be appropriately defined for data generated under different experimental conditions, and are not confined to data types represented only by vectors. A number of distance measures are readily available for several types of biological data, including DNA sequences \citep{wu1997measure}, genetic markers such as SNPs \citep{goh2011assessing}, gene expression data \citep{priness2007evaluation}, longitudinal gene expression data \citep{minas2011distance} and proteins \citep{hollich2005assessment}. Moreover, for a specific data type, more than one distance measure can be deployed to capture different aspects of the data. This can ultimately lead to the discovery of different types of associations.


Only a handful of statistical procedures are currently available to determine whether there is an association between paired distance matrices, and they all rely on computationally intensive procedures for statistical inference. The most commonly used and well-known procedure is the Mantel test \citep{mantel1967detection}, a generalization of Pearson's correlation. The test has already been found useful in many applications arising in bioinformatics \citep{shannon2002mantel,beckmann2005haplotype,sun2011identification}.
%
%
This testing procedure, however, has some limitations. First, when used to compare paired scalar-valued observations using the Euclidean distance, the test exhibits less power than a Pearson's correlation test \citep{peres2001well,legendre2010comparison}. Secondly, computationally intensive Monte Carlo permutations are required to assess significance. This procedure, which assumes exchangeability of the sampling units under the null hypothesis of no association, introduces sampling errors which leads to inaccurate estimation of small p-values. This is especially true when too few permutations are used, generally between  $O(10^3)$ and $O(10^5)$ \citep{berry1983moment,mielke2007permutation,knijnenburg2009fewer}. For example, in order to obtain a permutation p-value within $10^{-5}$ of the true p-value, it has been shown that $O(10^7)$ permutations are required, which is computationally infeasible without access to specialised high-performance computational resources. This is the case in all situations where thousands or even millions of tests are required, such as in eQTL mapping studies in which the entire genome is scanned in search of potential associations. In such large-scale applications, it is not uncommon to settle for a much smaller number of Monte Carlo permutations per test, which in turn can lead to inflated family-wise type I error rates \citep{phipson2010permutation}.



In this article we propose a novel test, the generalized RV (GRV) test, to detect association between paired distance matrices. This test is derived as a generalization of the classical multivariate RV test of no correlation between paired random vectors, originally proposed by \cite{escoufier1973traitement}. While inference can be performed via computationally intensive Monte Carlo permutations, we also derive an approximation to the exact null distribution which would be obtained by enumerating all permutations. Using the proposed null distribution, approximate p-values can be estimated in closed-form without the need for computationally intensive procedures. The proposed GRV test is introduced in Section \ref{grv_test}. In Section \ref{sims} we describe power studies showing that the GRV test can offer greater statistical power than that achieved by the Mantel test even when many millions of Monte Carlo permutations are performed. Section \ref{apps} presents an application of the GRV test in cancer research, where the interest lies in detecting biological pathways characterized by a non-random association between genetic and gene expression sample variability. Concluding remarks are presented in Section \ref{concs}.

\section{The generalized RV test}\label{grv_test}

\subsection{Problem statement}

Assume we have observed paired measurements $\{(x_i, y_i)\}_{i=1}^N$ on $N$ independent biological sampling units. Each one of the $\{x_i\}_{i=1}^N$ and $\{y_i\}_{i=1}^N$ measurement sets is represented by a particular data structure, such as continuous or discrete vectors, although the data do not necessarily need to be represented as vectors. For instance, the measurements might be structured as graphs or trees (e.g. representing biological networks or ontologies). We also suppose that suitably chosen semi-metric or metric distances $d_x$ and $d_y$ have been identified that capture the dissimilarity between any sample pair $\{(x_i, x_j)\}_{i\neq j}$ and $\{(y_i, y_j)\}_{i\neq j}$, respectively. On evaluating the distances for all possible pairs, we obtain the paired $N\times N$ distance matrices $\bm{\Delta}_{x}=\left\{d_x(x_i,x_j)\right\}_{i,j=1}^N$ and $\bm{\Delta}_{y}=\left\{d_y(y_i,y_j)\right\}_{i,j=1}^N$. 

Using the random sample, evidence of non-random association between these paired genomic measurements can be assessed by testing the null hypothesis that $\bm{\Delta}_{x}\neq a\bm{\Delta}_{y}$, for some positive constant $a$. This constant represents possible scaling differences between the elements of each distance matrix, which may arise because of the chosen distance measures. Under this null hypothesis, the relationship between the elements in $\bm{\Delta}_{x}$ is not maintained by the corresponding elements in $\bm{\Delta}_{y}$. The alternative hypothesis is that the distance matrices are equal up to a constant, in which case every distance $d_x(x_i,x_j)$ is linearly related to the paired distance $d_y(y_i,y_j)$. We are particularly concerned with settings where inferences must be drawn for a very large number of such tests simultaneously, hence the computational cost in obtaining a p-value for each test should be kept as low as possible.  


\subsection{The proposed test statistic}

When the observations are vectorial, the mean-centered vectors $\{\bm{x}_i\}_{i=1}^N \in \mathbb{R}^P$ and $\{\bm{y}_i\}_{i=1}^N \in \mathbb{R}^Q$ can be arranged in paired data matrices, $\bm{X}$ and $\bm{Y}$, of dimensions $N \times P$ and $N \times Q$, respectively. In this case, in order to establish whether the $P$ and $Q$ variables are correlated, the null hypothesis that $\bm{\Sigma}_{xy}=\bm{0}$ is commonly tested, where $\bm{\Sigma}_{xy}$ is the $P\times Q$ covariance matrix containing the true covariances between the $P$ variables observed in $\{\bm{x}_i\}_{i=1}^N$ and $Q$ variables observed in $\{\bm{y}_i\}_{i=1}^N$. The RV statistic has been proposed for this task and arises as a generalization of Pearson's correlation \citep{escoufier1973traitement}. 
As with Pearson's correlation, RV is computed as the ratio of covariance to the square-rooted product of the variances,
\begin{equation}\label{obs_rv_1}
\textrm{RV}(\bm{X},\bm{Y})=\frac{\tr\left(\bm{X}^T\bm{YY}^T\bm{X}\right)}{||\bm{X}^T\bm{X}||||\bm{Y}^T\bm{Y}||},
\end{equation}
where $\tr(\cdot)$ is the trace function and $||\bm{A}||=\sqrt{\tr(\bm{A}^T\bm{A})}$ denotes the Frobenius norm for matrix $\bm{A}$. The RV statistic ranges between $0$ and $1$, with no association when $\bm{X}^T\bm{Y}=\bm{0}$, and perfect association when there exists a linear mapping which relates every $P$-dimensional observation in $\bm{X}$ to every $Q$-dimensional observation in $\bm{Y}$. Thus larger values provide evidence against the null hypothesis. When $P=Q=1$, $\textrm{RV}(\bm{X},\bm{Y})=\textrm{cor}^2(\bm{X},\bm{Y})$, where cor$(\cdot,\cdot)$ denotes Pearson's correlation coefficient. 

  
We first observe that the RV test can be interpreted in terms of the Euclidean distance matrices $\bm{\Delta}_x$ and $\bm{\Delta}_y$ arising by applying the Euclidean distance measure to $\{\bm{x}_i\}_{i=1}^N$ and $\{\bm{y}_i\}_{i=1}^N$, respectively. This is because, due to properties of the trace operator, $\tr(\bm{X}^T\bm{Y}\bm{Y}^T\bm{X})=\tr(\bm{XX}^T\bm{YY})$, $||\bm{X}^T\bm{X}||=||\bm{XX}^T||$ and $||\bm{Y}^T\bm{Y}||=||\bm{YY}^T||$, so that RV statistic \eqref{obs_rv_1} can be written in equivalent form as
\begin{equation}\label{rv_2}
\textrm{RV}(\bm{X},\bm{Y})=\frac{\tr\left(\bm{XX}^T\bm{YY}^T\right)}{||\bm{XX}^T||||\bm{YY}^T||}.
\end{equation}
Note that \eqref{obs_rv_1} differs from \eqref{rv_2} in that emphasis is placed on the two symmetric $N\times N$ matrices $\bm{XX}^T$ and $\bm{YY}^T$, instead of the four covariance matrices $\bm{X}^T\bm{Y}\in\mathbb{R}^{P\times Q}$, $\bm{Y}^T\bm{X}\in\mathbb{R}^{Q\times P}$, $\bm{X}^T\bm{X}\in\mathbb{R}^{P\times P}$ and $\bm{Y}^T\bm{Y}\in\mathbb{R}^{Q\times Q}$. In settings where $P,Q>N$, which are common when comparing biological measurements, it is computationally convenient to use \eqref{rv_2} over \eqref{obs_rv_1}. 

In addition, the $N\times N$ matrices $\bm{XX}^T$ and $\bm{YY}^T$ can each be written in terms of the Euclidean distance matrices $\bm{\Delta}_{x}$ and $\bm{\Delta}_{y}$, respectively. This is due to the results that
\begin{equation}\label{mds_embed}
\bm{XX}^T=-\frac{1}{2}\bm{C}\bm{\Delta}^2_{x}\bm{C}\quad\textrm{and}\quad \bm{YY}^T=-\frac{1}{2}\bm{C}\bm{\Delta}^2_{y}\bm{C},\nonumber
\end{equation}
where $\bm{C}=\left(\bm{I}_N-\bm{J}_N/N\right)$ is the symmetric $N\times N$ centering matrix where $\bm{I}_N$ is the identity matrix of size $N$ and $\bm{J}_N$ is the square matrix of ones of size $N$ \citep{gower1966sdp,Borg2005}. Also, $-\frac{1}{2}\bm{C}\bm{\Delta}^2_{x}\bm{C}=\bm{G}_x$, where $\bm{G}_x$ is the square, symmetric and real-valued Gower's centered inner product matrix, and similarly for $\bm{G}_y$, so that $\tr\left(\bm{X}\bm{X}^T\bm{YY}^T\right)=\tr\left(\bm{G}_x\bm{G}_y\right)$, $||\bm{XX}^T||=||\bm{G}_x||$ and $||\bm{YY}^T||=||\bm{G}_y||$. Thus RV is completely specified by the elements of the Euclidean distance matrices.

Based on these observations, we propose a generalization for the RV statistic that is entirely specified by any pair of distance matrices, not necessarily Euclidean, in order to test the null hypothesis that $\bm{\Delta}_{x}\neq a\bm{\Delta}_{y}$. The generalized RV (GRV) test statistic is defined as
\begin{equation}\label{drv_stat}
\textrm{GRV}(\bm{G}_x,\bm{G}_y)=\frac{\tr\left(\bm{G}_x\bm{G}_y\right)}{\left|\left|\bm{G}_x\right|\right|\left|\left|\bm{G}_y\right|\right|},
\end{equation}
noting the implicit assumption $\left|\left|\bm{G}_x\right|\right|\left|\left|\bm{G}_y\right|\right|>0$ which is always satisfied in practice; $\left|\left|\bm{G}_x\right|\right|>0$ since $\bm{G}_x$ contains real-valued elements which are not all zero, and similarly for $\bm{G}_y$. 

It is insightful to understand in which ways the proposed GRV statistic is similar to, and differs from, the classical Mantel statistic. The Mantel statistic is computed as the correlation between the upper-triangular distances of each distance matrix. Denote by $\bm{v}_{x}$ the column vector containing the $A=N(N-1)/2$ upper-triangular distances $\{d_x(x_i,x_j)\}_{i>j}$, and similarly for $\bm{v}_{y}$. The Mantel statistic is then computed as $r_M(\bm{\Delta}_{x},\bm{\Delta}_{y})=\textrm{cor}\left(\bm{v}_{x},\bm{v}_{y}\right)$, i.e., \begin{equation}
r_M(\bm{\Delta}_{x},\bm{\Delta}_{y}) =\frac{1}{A-1}\sum_{i>j}\left(\frac{d_{x}(x_i,x_j)-\bar{x}}{s_x}\right)\left(\frac{d_{y}(y_i,y_j)-\bar{y}}{s_y}\right),\nonumber
\end{equation}
where $\bar{x}=\sum_{i>j}d_x(x_i,x_j)/A$, $s^2_x=\sum_{i>j}(d_x(x_i,x_j)-\bar{x})^2/(A-1)$, and similarly for $\bar{y}$ and $s^2_y$. The statistic therefore standardizes the upper-triangular elements of each distance matrix to have zero mean and unit variance. In this way the distances between paired distance matrices can be directly compared.  

It can be shown that, like Mantel, GRV is also a correlation coefficient. In particular, it can be proved that GRV is equal to the correlation between the vectorized matrices $\bm{G}_x/||\bm{G}_x||$ and $\bm{G}_y/||\bm{G}_y||$, i.e., the $N^2$-dimensional vectors $\bm{g}_x$ and $\bm{g}_y$ with elements $\{g_{x}(x_i,x_j)/||\bm{G}_x||\}_{i,j=1}^N$ and $\{g_{y}(y_i,y_j)/||\bm{G}_y||\}_{i,j=1}^N$, respectively. That is, $\textrm{GRV}(\bm{G}_x,\bm{G}_y)=\textrm{cor}\left(\bm{g}_x,\bm{g}_y\right)$  (see Appendix \ref{grv_cor} for a proof). While both Mantel and GRV are correlation coefficients, the difference between them lies in the methods of standardization applied to the distances in each case. In Mantel, the upper-triangular distances are subjected to a classical standardization, where their mean is subtracted and they are divided by their standard deviation. In GRV, however, all distance elements are considered, and they are squared, double-centered and normalized by dividing by their Frobenius norm. This difference leads to greater power of the GRV test to detect association between paired distance matrices than the Mantel test, as demonstrated in Section \ref{sims}. Further properties of the GRV test statistic are provided in Appendix \ref{properties}, including a discussion of how it overcomes the limitation of Mantel for scalar-valued observations and Euclidean distances.

\begin{sidewaystable}
\begin{center} 
\caption{Power of the GRV and Mantel tests at a significance level of $0.1\%$ for $N=\{30,50,70,90,100\}$ and all combinations of the genetic (gen) identity-by-state (IBS), Sokal and Sneath (SS), and Rogers and Tanimoto I (RTI) distances and gene expression (gex) Euclidean (Euc), Manhattan (Man) and Mahalanobis (Mah) distances. For the GRV test, the standard deviation of the power estimate is given in brackets. For the Mantel test, the confidence interval obtained with $95\%$ coverage probability is given in square brackets. The average number of Monte Carlo permutations required to achieve the power estimates for Mantel are stated below the confidence interval in millions.}
\label{table_grv_mantel}
\begin{tabular}{llllrlrlrlrlr}
 \toprule
Gen&Gex&Test& \multicolumn{10}{c}{$N$}\\
dist&dist&&\multicolumn{2}{c}{$30$}&\multicolumn{2}{c}{$50$}&\multicolumn{2}{c}{$70$}&\multicolumn{2}{c}{$90$}&\multicolumn{2}{c}{$100$}\\
\midrule
IBS&Euc&GRV&$0.126$&$(0.042)$&$0.351$&$(0.064)$&$0.566$&$(0.074)$&$0.712$&$(0.049)$ &$0.780$&$(0.051)$\\
&&Mantel&$0.081$&$[0.042,0.126]$&$0.116$&$[0.056,0.183]$&$0.225$&$[0.145,0.291]$&$0.275$&$[0.228,0.325]$&$0.322$&$[0.272,0.374]$\\
&&&&$3.081$&&$1.575$&&$9.921$&&$13.51$&&$19.61$\\
&Man&GRV&$0.139$&$(0.049)$&$0.375$&$(0.068)$&$0.590$&$(0.066)$&$0.730$&$(0.059) $&$0.772$&$(0.065)$\\
&&Mantel&$0.069$&$[0.023,0.120]$&$0.151$&$[0.091,0.226]$&$0.240$&$[0.173,0.306]$&$0.302$&$[0.238,0.355]$&$0.378$&$[0.314,0.444]$\\
&&&&$1.859$&&$2.134$&&$6.290$&&$13.80$&&$8.794$\\
&Mah&GRV&$0.394$&$(0.061)$&$0.846$&$(0.047)$&$0.952$&$(0.030)$&$0.978$&$(0.020) $&$0.990$&$(0.014)$\\
&&Mantel&$0.100$&$[0.049,0.169]$&$0.311$&$[0.262,0.355]$&$0.542$&$[0.513,0.572]$&$0.715$&$[0.694,0.735]$&$0.776$&$[0.761,0.789]$\\
&&&&$2.818$&&$28.53$&&$135.9$&&$424.2$&&$666.8$\\
\midrule
SS&Euc&GRV&$0.038$&$(0.029)$&$0.157$&$(0.042)$&$0.299$&$(0.067)$&$0.470$&$(0.064)$&$0.556$&$(0.063)$\\
&&Mantel&$0.035$&$[0.006,0.064]$&$0.111$&$[0.069,0.154]$&$0.180$&$[0.112,0.245]$&$0.268 $&$[0.202,0.329]$&$0.319$&$[0.254,0.381]$\\
&&&&$0.971$&&$5.318$&&$3.032$&&$7.445$&&$10.69$\\
&Man&GRV&$0.045$&$(0.028)$&$0.160$&$(0.046)$&$0.308$&$(0.068)$&$0.503$&$(0.073) $&$0.556$&$(0.068)$\\
&&Mantel&$0.047$&$[0.018,0.073]$&$0.121$&$[0.080,0.171]$&$0.195$&$[0.133,0.269]$&$0.310$&$[0.242,0.387]$&$0.363$&$[0.290,0.424]$\\
&&&&$2.879$&&$10.27$&&$4.371$&&$7.325$&&$9.232$\\
&Mah&GRV&$0.092$&$(0.041)$&$0.481$&$(0.074)$&$0.784$&$(0.059)$&$0.907 $&$(0.037)$&$0.927$&$(0.034)$\\
&&Mantel&$0.072$&$[0.030,0.112]$&$0.266$&$[0.194,0.340]$&$0.510$&$[0.448,0.564]$&$0.652$&$[0.613,0.687]$&$0.723$&$[0.689,0.756]$\\
&&&$$&$5.567$&$$&$8.635$&$$&$21.92$&$$&$74.12$&$$&$40.01$\\
\midrule
RTI&Euc&GRV&$0.044$&$(0.027)$&$0.159$&$(0.047)$&$0.340$&$(0.062)$&$0.551$&$(0.067)$&$0.654$&$(0.063)$\\
&&Mantel&$0.032$&$[0.010,0.065]$&$0.106$&$[0.061,0.154]$&$0.205$&$[0.142,0.265]$&$0.302$&$[0.233,0.366]$&$0.339$&$[0.272,0.397]$\\
&&&$$&$1.327$&$$&$5.141$&$$&$8.112$&$$&$1.421$&$$&$16.36$\\
&Man&GRV&$0.055$&$(0.035)$&$0.172$&$(0.052)$&$0.363$&$(0.067)$&$0.540$&$(0.076) $&$0.639$&$(0.072)$\\
&&Mantel&$0.051$&$[0.019,0.089]$&$0.124$&$[0.067,0.171]$&$0.216$&$[0.154,0.288]$&$0.360$&$[0.282,0.432]$&$0.401$&$[0.321,0.465]$\\
 &&&&$1.373$&&$4.512$&&$4.371$&&$9.289$&&$7.233$\\
&Mah&GRV&$0.173$&$(0.055)$&$0.675$&$(0.068)$&$0.925$&$(0.038)$&$0.987$&$(0.015)$&$0.993$&$(0.010)$\\
&&Mantel&$0.062$&$[0.011,0.119]$&$0.241$&$[0.180,0.314]$&$0.515$&$[0.478,0.554]$&$0.688$&$[0.672,0.702]$&$0.765$&$[0.755,0.776]$\\
&&&$$&$0.8066$&$$&$11.89$&$$&$151.0$&$$&$654.4$&$$&$1068$\\
\bottomrule
\end{tabular}

\end{center}
\end{sidewaystable}

\subsection{Approximate null distribution}
 
Under the null hypothesis, the sampling distribution of GRV is generally unknown. This is because the quantity $T=\tr(\bm{G}_x\bm{G}_y)$ in the numerator is completely specified by the elements of $\bm{\Delta}_{x}$ and $\bm{\Delta}_{y}$, which have unknown distributions and are correlated. The null sampling distribution can be generated by using permutations of one of the centered inner product matrices, $\bm{G}_y$, say. For each of $N_{\pi}$ permutations $\pi\in \Pi$, which are one-to-one mappings of the set $\{1,\ldots,N\}$ to itself, the rows and columns of $\bm{G}_y$ are simultaneously permuted by $\pi$ and denoted $\bm{G}_{y,\pi}$. This generates the set $\{\hat{\textrm{GRV}}(\bm{G}_x,\bm{G}_{y,\pi})\}_{\pi\in\Pi}$ and the p-value of an observed GRV statistic, $\hat{\textrm{GRV}}(\bm{G}_x,\bm{G}_y)$, can be obtained as the proportion of permuted statistic values greater than or equal to $\hat{\textrm{GRV}}(\bm{G}_x,\bm{G}_y)$. This is a right-tailed test as larger GRV values provide evidence against the null.

In order to estimate the p-value without permutations, we adopt a moment matching approach where the exact null distribution which would be obtained if all $N!$ permutations were used is approximated by a continuous distribution. In particular, we approximate the null distribution by the Pearson type III distribution, which has been shown to capture skewed characteristics of null sampling distributions \citep{mielke2007permutation, josse2008testing}. To use this distribution we require the first few moments of the exact permutation distribution of $T$; the mean is $\mu=\frac{1}{N!}\sum_{\pi\in\Pi}\hat{T}_{\pi}$, the variance is 
$\sigma^2=\frac{1}{N!}\sum_{\pi\in\Pi}\hat{T}^2_{\pi}-{\mu}^2$ and the skewness is 
\begin{equation}
\gamma=\frac{\frac{1}{N!}\sum_{\pi\in\Pi}\hat{T}^3_{\pi}-3\mu\sigma^2-\mu^3}{\sigma^3},\nonumber
\end{equation}
where $\hat{T}_{\pi}=\tr(\bm{G}_x\bm{G}_{y,\pi})$ and $\Pi$ contains all $N!$ permutations. No permutations are needed to compute $\{\mu,\sigma,\gamma\}$, since closed-form expressions of these quantities are retrievable via the analytical results of \cite{kazi1995refined}. These results require that $\bm{G}_x$ and $\bm{G}_y$ are square, symmetric and centered; properties satisfied by definition. 

On obtaining closed-form expressions for the mean, variance and skewness of the exact permutation distribution of $T$, we standardize $T$ by subtracting $\mu$ and dividing by $\sigma$. The resulting $T_s=(T-\mu)/\sigma$ is then assumed to have the Pearson type III distribution under the null with probability density function (PDF) denoted $f_{T_s}(t;\gamma)$; see \cite{mielke2007permutation} for the full definition of the distribution. We denote the CDF of $T_s$ by $\mathcal{F}_{T_s}(t;\gamma)$. The approximate null distribution of GRV$(\bm{G}_x,\bm{G}_y)=T/||\bm{G}_x||||\bm{G}_y||$ can then be derived by a simple transformation of the distribution of $T_s$. The CDF of GRV, $\mathcal{F}_{\text {GRV}}(x;\gamma)$, is given by
 
\begin{equation}
\mathcal{F}_{\textrm{GRV}}(x;\gamma)=\mathcal{F}_{T_s}\left(\frac{x||\bm{G}_{x}||||\bm{G}_{y}|| - \mu}{\sigma};\gamma\right),\nonumber
\end{equation}
and this is a valid CDF since $\mathcal{F}_{T_s}(;\gamma)$ is a valid CDF. The p-value of an observed GRV statistic $\hat{\textrm{GRV}}(\bm{G}_x,\bm{G}_y)$ is then estimated by $1-\mathcal{F}_{\text {GRV}}(\hat{\textrm{GRV}}(\bm{G}_x,\bm{G}_y);\gamma)$. The PDF $f_{\textrm{GRV}}(x,\gamma)$ is found by differentiation as 
\begin{equation} 
f_{\textrm{GRV}}(x,\gamma)=\left(\frac{||\bm{G}_{x}||||\bm{G}_{y}||}{\sigma}\right)f_{T_s}\left(\frac{x||\bm{G}_{x}||||\bm{G}_{y}|| - \mu}{\sigma};\gamma\right),\nonumber
\end{equation} 
since $||\bm{G}_{x}||||\bm{G}_{y}||>0$ and $\sigma>0$. 

\section{Power Studies}\label{sims}

In this section we assess the power of the proposed GRV test and compare it to the classical Mantel test. A Monte Carlo procedure is used to estimate the power of the GRV test, and in order to estimate the power of the Mantel test we use the recently proposed algorithm of \cite{gandy2011algorithm}. This algorithm estimates the power of a Monte Carlo testing procedure and states a confidence interval around this estimate boasting a guaranteed coverage probability. It also states the average number of Monte Carlo permutations required to achieve the given power estimate. It requires specification of the maximum length of the resulting confidence interval and the desired coverage probability, and then runs as many permutations as required to yield a power estimate with a confidence interval of length no greater than specified. Using this approach we are able to demonstrate that, at least for the experimental setups considered, many millions of permutations are needed for the Mantel test to achieve power of the same precision as the GRV test, and that the GRV test achieves greater power. 

We use simulated data that mimics the experimental data of a typical eQTL application where transcriptional measurements are paired with SNPs. 
We first generate an $N\times P$ matrix $\bm{X}$
containing $N$ simulated SNPs (i.e., minor allele counts), denoted $\bm{x}_i=(x_{i1},\ldots,x_{iP})^T$ for $i=1,\ldots,N$, with varying minor allele frequencies (MAFs) at each of the $P$ SNPs. The MAF of SNP $p$, $m_p$, is first simulated from a uniform distribution $U(0.1,0.5)$, and the allele count is then simulated from a multinomial distribution with probabilities $(1-m_p)^2$, $2m_p(1-m_p)$, and $m_p^2$ of observing $0$, $1$ and $2$ minor alleles, respectively.  The paired data matrix $\bm{Y}=(\bm{y}_1,\ldots,\bm{y}_{N})^T$ is then generated as follows. The $N\times 1$ vector $\bm{z}=\bm{X}\bm{1}_P=(z_1,\ldots,z_N)^T$ containing the row sums of $\bm{X}$, i.e., the minor allele counts across the $P$ SNPs, is computed, and $\bm{y}_i=z_i\bm{1}_Q+\bm{e}_i$ for $i=1,\ldots,N$, where $\bm{e}_i\sim N_Q(\bm{\mu},\bm{\Sigma})$, with $\bm{\mu}=(\mu_1,\ldots,\mu_Q)^T$ and $\mu_q\sim U(0,1)$ for $q=1,\ldots,Q$, and $\bm{\Sigma}$ a random $Q\times Q$ Wishart matrix. We then consider the widely-used identity-by-state (IBS) distance for the simulated genetic markers in $\bm{X}$, which measures the degree to which alleles are shared across the SNPs (see, for instance, \cite{goh2011assessing}), in addition to the Sokal and Sneath, and Rogers and Tanimoto I distances \citep{selinski2005cluster}. For the observations in $\bm{Y}$ we use the Euclidean, Manhattan and Mahalanobis distances. See Appendix \ref{dists} for details on these distances.

For $P=2$, $Q=10$, and each of $N=\{30,50,70,90,100\}$, the power of the GRV test is computed using $50$ Monte Carlo runs, and each time generating $50$ paired datasets as above. For each genetic and gene expression distance, the GRV test is applied and the proportion of p-values less than or equal to the significance level of $0.1\%$ is recorded. The mean power across all $50$ Monte Carlo runs for each distance combination and value of $N$ is reported in Table \ref{table_grv_mantel}, in addition to the standard deviation of these estimates. As expected, the power increases with $N$.

We then estimate the power of the Mantel test for each of the above settings using the algorithm of \cite{gandy2011algorithm}. For each setting we bound the confidence interval length by twice the standard deviation of the corresponding estimate for the GRV test. That is, we consider one standard deviation on either side of the power estimate as an empirical indication of the precision achieved by the GRV test. We monitor the number of Monte Carlo permutations required to obtain power estimates with such confidence intervals with a coverage probability of $95\%$. These results are also given in Table $1$, where the power is stated alongside the confidence interval, and the number of Monte Carlo permutations required is stated on a separate line below the confidence interval. 

In all settings the GRV test achieves greater power than the Mantel test, even when using large numbers of Monte Carlo permutations. Note that while the Mantel power estimates improve with $N$, as expected, the number of Monte Carlo permutations varies between $O(10^5)$ to $O(10^{9})$ non-linearly with $N$. This is an artefact of the algorithm; the expected number of permutations depends on the length of the confidence interval sought and the region of the confidence interval in $[0,1]$ (see \cite{gandy2011algorithm} for more details). An example of how the standardization used by GRV yields greater power than that used by Mantel is presented in Appendix \ref{pows}, in addition to simulation results demonstrating that the GRV test attains the nominal significance level of a given test. 

\section{Application to ovarian cancer}\label{apps}

Around $225,000$ women were estimated to have developed ovarian cancer in 2008, of which over $80\%$ are Epithelial Ovarian Cancer (EOC). In the UK EOC has particularly poor prognosis with only $34\%$ of patients surviving 5 years after diagnosis.

In many cancers, the ability to measure some biomarker that can predict the likely response to a chemotherapeutic agent could prevent patients unnecessarily suffering side-effects from ineffective treatments, and may suggest the best course of action where multiple treatments are available. Unfortunately, while high-throughput gene expression profiling has been successful in identifying prognostic `signatures' for many cancers \citep{vantveer08}, solid tumour gene expression profiles have practical drawbacks preventing their use as prognostic or predictive biomarkers in the clinic, especially in terms of obtaining mRNA from the tumour and the transitive nature of gene expression \citep{sawyers08}. However, circulating tumour DNA can be isolated from serum samples in ovarian cancer patients \citep{hickey99}, which reflect an attractive source for genetic biomarkers both because of the relative ease of accessibility of the patient material and the long-term stability of the genetic sequence. Additionally, treating biological pathways as multidimensional biomarkers may provide substantial improvements in terms of robustness over single agents \citep{dai11}.

We sought to discover if it might be possible to use genetic biomarkers as surrogates for the transcriptional activation of pathways. Such pathway-based genetic biomarkers may suggest starting points for the development of clinical diagnostic tests to predict the outcome of a therapeutic exploiting a related pathway.

In order to identify pathways for which the profile of SNP genotypes across genes in the pathway closely reflects the state of expression of the genes in the pathway, we applied the GRV test to evaluate the significance of the degree of similarity in distance matrices derived from SNP genotypes and gene expression levels from a large set of selected pathways, for two independent cohorts. 

\subsection{Datasets}

For this study we use two independent cohorts of ovarian cancer patients: one collected by the TCGA consortium, and an independent cohort for validation. For the TCGA cohort, raw data CEL files from Affymetrix SNP $6.0$ and from Affymetrix HT-HGU133A microarray profiling $499$ primary high grade serous ovarian tumours were obtained from the TCGA web repository. For the Affymetrix SNP 6.0 arrays, normalization and genotype calling were performed using the CRLMM method \citep{carvalho2007}. Current annotations were obtained from Affymetrix's NetAffx Analysis Center
and SNPs were annotated with their associated Gene Symbols. Once SNP probes were mapped to Gene Symbols, lists of SNPs corresponding to each CPDB pathway were obtained by selecting all SNP probes mapping to Gene Symbols that were included in the pathway's list of \texttt{hgnc\_symbol\_ids} as downloaded from ConsensusPathDB (CPDB) \citep{cpdb}. For the Affymetrix HT-HGU133A arrays, data normalization was performed using the RMA method \citep{irizarry03}. Probe-set mapping to Gene Symbols was performed using the \textit{hthgu133a.db} package from Bioconductor. Once probe-sets were mapped to Gene Symbols, lists of probe-sets corresponding to each CPDB pathway were obtained by selecting all probe-sets mapping to Gene Symbols that were included in the pathway's list of \texttt{hgnc\_symbol\_ids} as downloaded from CPDB.
     
As an independent validation data set, raw data CEL files from both Affymetrix SNP $6.0$ and Affymetrix HGU133plus2 microarrays profiling $60$ primary high grade serous ovarian tumours were obtained from the Hammersmith Biobank Resource Centre. The datasets were provided courtesy of the Genome Institute of Singapore. Raw data for this independent tumour cohort was processed identically to those obtained from TCGA, although mapping of gene expression probe-sets to Gene Symbols was performed using the \textit{hgu133plus2.db} package from Bioconductor. The tables of genotype calls and normalized gene expression values, mapped to pathways which were used in this analysis, are provided in Supplementary Tables S1 and S2, respectively.

We wanted to reduce the possibility that any identified associations simply reflected genome-wide similarity structures between the patients (e.g. due to molecular subtypes of disease) or involved so few measurements that they weren't indicative of the entire pathway. From the full list of $4,119$ CPDB pathways we filtered out all those with less than 5 or more than 200 probe sets in either gene expression or SNP genotype array datasets. A total of $479$ pathways remained after this filtering, and these pathways are listed in Supplementary Table S3 along with their mappings to probes in each of the two datasets.

\subsection{Experimental results}

Since different distance measures highlight different relationships among genetic markers and features of gene expression data \citep{priness2007evaluation,song2012}, we sought to perform a meta-analysis to combine the results of similarity testing across a range of distance measures for both the genotypes and gene expression values. The rationale being that if a given pathway shows similar distance matrices at the genotype and gene expression level for a number of different distance measures, the observed similarity is unlikely to be due to a quirk of one particular distance measure. For each of the $479$ pathways, sample-wise distance matrices were calculated using $8$ different distance measures (Euclidean, Mahalanobis, Manhattan, Maximum, Bray-Curtis, Pearson's correlation, Spearman's correlation and Cosine angle distances) for the gene expression data and $5$ distance measures (IBS, SS, RTI, Simple Matching and Hamman I) for the SNP genotypes; see Appendix \ref{dists} for details on these distances. The association between the paired distance matrices, one derived from the genotypes of SNPs mapping to genes in the pathway and the other derived from the expression levels of genes in the pathway, was evaluated using the GRV test. This resulted in $40$ individual p-values for each pathway, one for each combination of distance measures used. An illustration of how the the null distribution compares with the permutation distribution when applied to this cancer data is presented in Appendix \ref{data}. The $40$ individual p-values for each pathway were then combined using the `maximum' method as implemented in the R package \textit{MetaDE}. The combined significance estimates from running this meta-analysis for each pathway are given in Supplementary Table S4. This meta-analysis was then repeated for the validation cohort. 

\begin{figure}[h!]
        \center
        \includegraphics[scale=0.6]{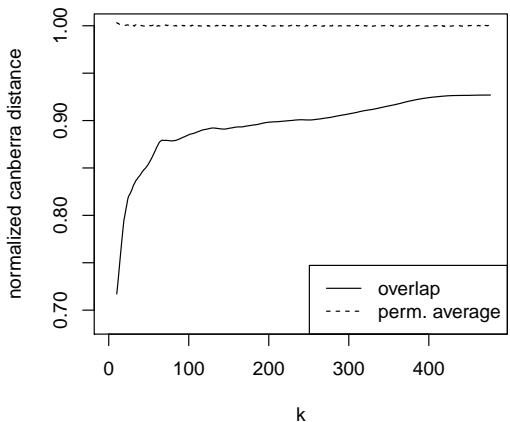}
        \caption{Plot of the normalized Canberra distance against $k$ for the top $k$ elements in two lists of pathways. Each pathway list is ranked according to meta-analysis estimate of significance of association between genotype and gene expression over $40$ combinations of distance measures, one ranked list per cohort. The dotted line represents the average normalized Canberra distance between the TCGA data-derived pathway list and each of $5,000$ random permutations of the ranks in the second list.}
\label{norm_can}
 \end{figure}

To obtain a perspective of the robustness/repeatability of the associations detected between genotype and gene expression at the pathway level across the two independent cohorts, we compared the ranked lists of pathways arising from the meta-analysis of each cohort. The degree of agreement between the two ranked lists was evaluated using the normalized Canberra distance \citep{jurman2008}. The general principle is to quantify the distance between the rankings of the top $k$ terms of two ranked lists, with each term's influence on the overall distance weighted by its position in its respective list. By evaluating the normalized Canberra distance for the top-ranked $k$ pathways, the relationship between overlap and ranking can be explored: smaller values indicate a lesser discrepancy in rankings among the top $k$ terms and therefore more similar rankings in the two lists. Figure \ref{norm_can} shows the normalized Canberra distance plotted against $k$.

\noindent It can be seen that the overlap of the meta-analysis results from the two cohorts is greatest for small $k$, implying that the highest-ranking pathways in each list are more similar than the lower-ranking pathways. Furthermore, for each $k$ we used $5,000$ permutations of the ranks in the second list to obtain a p-value estimate for the normalized Canberra distance values observed between the two rankings. These p-values across the range of $k$ were less than $0.005$ after false discovery rate multiple-testing adjustment, reflecting the fact that the overlap between the two ranked lists of pathways was significantly greater than that expected by chance. These results suggest that the association between distance matrices based on the levels of expression of genes within certain pathways and distance matrices based on SNP genotypes associated with genes in the corresponding pathways is robust and reliably evaluated by the GRV test. The meta-analysis of the GRV test result across multiple distance measures demonstrates that these associations are not a result of peculiarities of particular distance measures, and certain pathways are shown to exhibit a very strong association in both cohorts. 

A particularly striking result of this study is the CPDB pathway `5-aminoimidazole ribonucleotide biosynthesis', which is ranked $\#1$ in the meta-analysis of GRV test results from the TCGA dataset and ranked $\#2$ in the meta-analysis of GRV test results from the validation dataset. The 5-aminoimidazole ribonucleotide is a precursor to de novo purine biosynthesis, which is essential for rapidy proliferating tissue \citep{firestine1993}. Pharmacological inhibition of de novo purine biosynthesis has proved to be a successful strategy in the clinic for treating cancers and inflammatory disorders \citep{christopherson2002}. It has been shown through a clinical trial that genetic polymorphisms could be biomarkers to predict response to methotrexate therapy, which inhibits de novo purine biosynthesis, in patients with rheumatoid arthritis \citep{dervieux2004}. The `methotrexate pathway' itself showed significant genotype-expression association in the TCGA cohort (ranked $\#17$), but not in the validation cohort (ranked $\#316$). The `5-aminoimidazole ribonucleotide biosynthesis' pathway may therefore suggest additional or alternative SNPs to use as genetic biomarkers for predicting outcome of inhibiting de novo purine biosynthesis.

Another result of note is that the CPDB pathway Platinum Pathway, Pharmacokinetics/Pharmacodynamics yields highly significantly similar tumour-wise distance matrices across all $40$ applications of the GRV test on this paired ovarian cancer data, for both cohorts investigated (ranked $\#14$ from the TCGA dataset and $\#42$ from the validation dataset). This is particularly interesting because platinum-based chemotherapy is the standard treatment for ovarian cancer, but unlike for many other solid cancers, survival rates in ovarian cancer have not improved noticeably in the $30$ years since this treatment was introduced \citep{vaughan11}. The principal reason for this is the high frequency of platinum-resistant relapse, which may occur through multiple mechanisms \citep{agarwal03,stronach11}. Therefore, being able to predict the response to the platinum-based DNA damaging agents could be of great clinical value in the treatment of ovarian cancer. 
Additionally, all genotype-derived and expression-derived distance matrices for the Base Excision Repair pathway are highly significantly similar in the TCGA cohort. This is of interest as the Base Excision Repair pathway is involved in repairing platinum-induced DNA damage. In fact, effective inhibition of this pathway (via PARP) in patients with BRCA1/2 mutations is associated with sensitivity to platinum chemotherapy \citep{fong2010}. 

%

\section{Conclusion}\label{concs}

In this work we have presented the GRV test as a novel procedure to detect association between paired distance matrices, which is applicable in any setting where suitable distance measures can be defined. Similarly to the widely used Mantel test, the GRV test can be directly applied whether the distance measures are metric or semi-metric, but overcomes the limitation of the Mantel test that, where paired data is vector-valued, a hypothesis of no correlation between the vectors of interest can be tested. As a further advantage of the GRV test over existing distance-based tests, an approximate null distribution is proposed such that inference can be performed without expensive Monte Carlo permutation procedures. Through extensive simulations we have demonstrated that the GRV test achieves greater power than the popular Mantel test. This greater power is exhibited even when many millions of permutations are used for the Mantel test, thus demonstrating its suitability as a tool for use in multiple-testing settings were many tests need to be simultaneously performed. 

The GRV test was applied to paired ovarian cancer datasets in which genome-wide profiles of SNP genotypes and gene expression levels were available for each patient across two independent cohorts ($N=499$ and $N=60$). The paired datasets were split into subsets in which the features all mapped to the same pathway (using mappings from the Consensus Pathway DB), so that each GRV test result would indicate the statistical significance of the similarity between the (tumour-wise) distance matrix derived from the genotype data and the distance matrix derived from the gene expression data. For each cohort a meta-analysis was performed across the GRV test results from $40$ different combinations of distance measures, and the pathways were ranked according to their significance across all distance measures. Permutation-based estimates of the significance of the overlap between the ranked lists of pathways from the two independent cohorts revealed a highly significant agreement in the two rankings, indicating a degree of robustness in the results of our analysis.

This application of the GRV test has demonstrated that the transcriptional activity of a number of pathways seem to be reliably predicted by sequencing a limited set of genetic markers that could be detected in circulating tumour DNA obtained from patients' serum. This relationship was particularly pronounced for the pathway `5-aminoimidazole ribonucleotide biosynthesis', for which the genotypes and the gene expression patterns were strikingly associated in two independent cohorts of patients. The association between genotypes and pathway-level gene expression profile is of particular interest in the case of the platinum response pathway, which is clearly clinically important in EOC given that the only standard treatment involves platinum-based chemotherapy. Following further investigation, these observations could potentially lead to development of non-invasive biomarkers that could predict patient response to front-line chemotherapy, and serve to stratify patients for selection into clinical trials involving alternative treatments.

As individual distance measures can yield different results in the analysis of associations across gene expression patterns, unless a single distance measure is selected {\it a priori} as the most appropriate for each dataset, combining results of multiple tests of association using different distance measures may provide more reliable results than a single test. A key advantage of the GRV test in our application to multivariate eQTL mapping of pathways in ovarian cancer is that it enables a meta-analysis of the associations between genetic distance and gene expression distance for different combinations of distance measures. Having a test for these associations that does not require expensive permutations enables a fast estimation of the robustness of the genotype-expression pathway level associations against quirks of particular definitions of distance. Given the robust associations discovered in this analysis, we suggest that it may be possible to predict the mode of activation of a pathway (in terms of which members are transcriptionally activated) in cancer patients, based on the genotype of the tumours' DNA across a limited panel of SNPs.

\section*{Acknowledgements}
The authors would like to thank Professor Edison T. Liu, Genome Institute of Singapore, and the Agency for Science Technology and Research of Singapore for use of ovarian cancer microarray data. E. C. was funded by Ovarian Cancer Action, and G. M. and C. M. acknowledge financial support from the EPSRC.

\appendix

\section{GRV statistic is a correlation coefficient}\label{grv_cor}

Consider the quantity cor$(\bm{g}_{x},\bm{g}_{y})$. To compute this we require the mean and standard deviation of the values in $\bm{g}_{x}$ and $\bm{g}_{y}$. The means are $0$ since $\bm{G}_{x}$ and $\bm{G}_{y}$ are centered matrices. The standard deviation of the elements in $\bm{g}_{x}$ is given by 
\begin{eqnarray}
\sqrt{\frac{1}{N^2 -1}\sum_{i=1}^N\sum_{j=1}^N\left(\frac{g_{x}(x_i,x_j)}{||\bm{G}_{x}||}\right)^2}&=&\sqrt{\frac{1}{||\bm{G}_{x}||^2\left(N^2 -1\right)}\sum_{i=1}^N\sum_{j=1}^N g^2_{x}(x_i,x_j)}\nonumber\\
&=&\sqrt{\frac{||\bm{G}_{x}||^2}{||\bm{G}_{x}||^2\left(N^2 -1\right)}}\nonumber\\
&=&\sqrt{\frac{1}{N^2 -1}},\nonumber
\end{eqnarray}
and similarly for $\bm{g}_{y}$. Thus the correlation of interest is given by
\begin{eqnarray}
\textrm{cor}\left(\bm{g}_{x},\bm{g}_{y}\right)&=&\frac{1}{N^2 - 1}\sum_{i=1}^N\sum_{j=1}^N \left(\frac{g_{x}(x_i,x_j)/|\bm{G}_{x}||}{\sqrt{\frac{1}{N^2-1}}}\right)\left(\frac{g_{y}(y_i,y_j)/||\bm{G}_{y}||}{\sqrt{\frac{1}{N^2-1}}}\right)\nonumber\\
&=&\frac{1}{N^2 - 1}\frac{1}{\left(\frac{1}{N^2-1}\right)}\sum_{i=1}^N\sum_{j=1}^N\frac{g_{x}(x_i,x_j)g_{y}(y_i,y_j)}{||\bm{G}_{x}||||\bm{G}_{y}||}\nonumber\\
&=&\frac{1}{||\bm{G}_{x}||||\bm{G}_{y}||}\sum_{i=1}^N\sum_{j=1}^N g_{x}(x_i,x_j)g_{y}(y_i,y_j)\nonumber\\
&=&\frac{\tr\left(\bm{G}_{x}\bm{G}_{y}\right)}{\left|\left|\bm{G}_{x}\right|\right|\left|\left|\bm{G}_{y}\right|\right|}\nonumber\\
&=&\textrm{GRV}(\bm{G}_{x},\bm{G}_{y}),\nonumber
\end{eqnarray}
as required.

\section{Properties of the GRV statistic}\label{properties}

The interpretation of GRV as a correlation coefficient indicates that it may range between $-1$ and $1$, with negative values indicating association in the form of a linear correlation of different sign (as with Pearson's correlation and Mantel). In this section we discuss how to interpret the values of GRV. 

To begin, we show that the GRV statistic is related to the Frobenius distance between $\bm{G}_x$ and $\bm{G}_y$. Recall that the Frobenius distance between two matrices $\bm{A}$ and $\bm{B}$ of equal dimensions is defined by $d_F(\bm{A},\bm{B})=\sqrt{\tr\left((\bm{A}-\bm{B})^T(\bm{A}-\bm{B})\right)}$. It is easily shown that the Frobenius distance between the scale invariant configurations $\bm{G}_x/||\bm{G}_x||$ and $\bm{G}_y/||\bm{G}_y||$ is related to the GRV statistic by
\begin{equation}\label{drv_dist}
d_{F}\left(\frac{\bm{G}_x}{||\bm{G}_x||},\frac{\bm{G}_y}{||\bm{G}_y||}\right)=\sqrt{2\left(1-\textrm{GRV}(\bm{G}_x,\bm{G}_y)\right)}.
\end{equation}
When $d_{F}\left(\bm{G}_x/||\bm{G}_x||,\bm{G}_y/||\bm{G}_y||\right)=0$, observe that perfect association is achieved when $\textrm{GRV}(\bm{G}_x,\bm{G}_y)=1$, i.e., when
\begin{equation}\label{perfect_asscc_eq}
\frac{\bm{G}_x}{||\bm{G}_x||}=\frac{\bm{G}_y}{||\bm{G}_y||}.
\end{equation} 
\noindent This equality, however, can only be attained if $\bm{G}_x$ and $\bm{G}_y$ are both positive semi-definite (having non-negative diagonals), or both indefinite (having non-negative and negative values on the diagonals). These occur if $d_x$ and $d_y$ are both metric, or semi-metric, respectively. When one distance function is metric and the other is semi-metric, perfect association cannot be attained, as the diagonals of $\bm{G}_x$ and $\bm{G}_y$ cannot be equal. In addition, negative values are attained for GRV for certain combinations of these metric properties. 

To see this, consider the upper and lower bounds of the GRV statistic, found as follows. Let the ordered eigenvalues of $\bm{G}_{x}$ and $\bm{G}_{y}$ be denoted $\{\lambda_{x,i}\}_{i=1}^N$ and $\{\lambda_{y,i}\}_{i=1}^N$, respectively, and consider the bounds of the quantity $\tr\left(\bm{G}_x\bm{G}_y\right)$. Using the result of \cite{lasserre1995trace}, which gives bounds for the trace of the product of two Hermitian matrices (square, complex-valued, and equal to their conjugate transpose), these are given by
\begin{equation}
\sum_{i=1}^N \lambda_{x,i}\lambda_{y,N-i+1}\leq \tr(\bm{G}_x\bm{G}_y)\leq \sum_{i=1}^N  \lambda_{x,i}\lambda_{y,i}.\nonumber
\end{equation}
The bounds for GRV are then given by  
\begin{equation}
\frac{ \sum_{i=1}^N \lambda_{x,i}\lambda_{y,N-i+1}}{||\bm{G}_x||||\bm{G}_y||}\leq \textrm{GRV}(\bm{G}_x,\bm{G}_y)\leq \frac{\sum_{i=1}^N \lambda_{x,i}\lambda_{y,i}}{||\bm{G}_{x}||||\bm{G}_{y}||},\nonumber
\end{equation}
 since $||\bm{G}_x||||\bm{G}_y||>0$. The numerators of the bounds of GRV are sums of eigenvalue products, and so may be non-negative or negative depending on the sign of the eigenvalues. This in turn depends on the distance functions satisfying the metric property; they are either both metric or both semi-metric, or one is metric and the other is semi-metric. Note that greater Frobenius distance values occur with lower GRV values which may be negative, and so the paired distance matrices are considered to be less associated.


\subsection{Both metric or semi-metric distances}

When the distances are both metric, $\bm{G}_x$ and $\bm{G}_y$ are positive semi-definite and the ordered eigenvalues $\{\lambda_{x,i}\}_{i=1}^N$ and $\{\lambda_{y,i}\}_{i=1}^N$ are non-negative \citep{krzanowski2000pma}. The summation in the lower bound contains the terms $\{\lambda_{x,i}\lambda_{y,N-i+1}\}_{i=1}^N$, which are therefore non-negative, so that GRV is non-negative. The minimum value of $0$ indicating no association is attained when $\tr\left(\bm{G}_x\bm{G}_y\right)=0$. This can be interpreted in terms of the $N\times N$ principal coordinate matrices arising from $\bm{\Delta}_{x}$ and $\bm{\Delta}_{y}$, denoted $\tilde{\bm{X}}$ and $\tilde{\bm{Y}}$, respectively. These matrices are found by classical multidimensional scaling (MDS) such that the rows are $N$-dimensional representations of the original observations in Euclidean space, and their pairwise Euclidean distances equal the corresponding pairwise distances in $\bm{\Delta}_{x}$ and $\bm{\Delta}_{y}$ \citep{torgerson1952multidimensional, gower1966sdp}. It then follows that $\bm{G}_x=\tilde{\bm{X}}\tilde{\bm{X}}^T$ and $\bm{G}_y=\tilde{\bm{Y}}\tilde{\bm{Y}}^T$, so that $\tr\left(\bm{G}_x\bm{G}_y\right)=\tr(\tilde{\bm{X}}\tilde{\bm{X}}^T\tilde{\bm{Y}}\tilde{\bm{Y}}^T)=0\Rightarrow \tilde{\bm{X}}^T\tilde{\bm{Y}}=\bm{0}$, i.e., the principal coordinate matrices are orthogonal. The maximum value GRV can take is $1$, since $\bm{G}_x$ and $\bm{G}_y$ have positive diagonal elements so that equality \eqref{perfect_asscc_eq} can be attained. In this case the distance matrices are equal up to a positive scaling factor, and there is perfect association. 

Note also that when $\bm{X}$ and $\bm{Y}$ are centered vector-valued observations with $\bm{X}^T\bm{Y}=\bm{0}$, and $d_x$ and $d_y$ are the Euclidean distance functions, GRV yields a value of $0$. This is because $\tilde{\bm{X}}\tilde{\bm{X}}^T=\bm{XX}^T$ and $\tilde{\bm{Y}}\tilde{\bm{Y}}^T=\bm{YY}^T$, so that $\tr\left(\bm{G}_x\bm{G}_y\right)=\tr(\tilde{\bm{X}}\tilde{\bm{X}}^T\tilde{\bm{Y}}\tilde{\bm{Y}}^T)=\tr\left(\bm{XX}^T\bm{YY}^T\right)$. This result applies when the data is scalar-valued, so GRV equals 0 inline with the raw data being uncorrelated. Thus GRV overcomes the limitation of Mantel.

On the other hand, when the distances functions are both semi-metric, the ordered eigenvalues $\{\lambda_{x,i}\}_{i=1}^N$ and $\{\lambda_{y,i}\}_{i=1}^N$ are both non-negative and negative, so that GRV may attain negative values. Since the diagonal elements of $\bm{G}_x$ and $\bm{G}_y$ are both positive and negative, there may exist two such matrices with equal diagonals. Hence there may exist a scenario in which equality \eqref{perfect_asscc_eq} is attained, and perfect association is indicated by a GRV value of $1$.

\subsection{Metric and semi-metric distances}

When one of the distances, $d_x$ for instance, is metric while $d_y$ is semi-metric, then only the ordered eigenvalues $\{\lambda_{x,i}\}_{i=1}^N$ are strictly non-negative, so that the summation in the lower bound may be negative. In this case, GRV may attain negative values. The maximum attainable value of GRV is not $1$ in this case, since equality \eqref{perfect_asscc_eq} cannot be attained. This is because the diagonals of $\bm{G}_x$ are positive while the diagonals of $\bm{G}_y$ are both positive and negative. Thus perfect association cannot be attained, but larger values indicate greater association. 

This result is not surprising because the relationship between pairwise distances with respect to $d_x$ and $d_y$ cannot be the same. Recall that metric distance functions satisfy the triangle inequality so that $d_x(x_i,x_j)\leq d_x(x_i,x_k)+d_x(x_k,x_j)$  for any three observations $x_i,x_j,x_k$ \citep{mardia1979bi}. The corresponding distances with respect to $d_y$ will not share this property as $d_y$ is semi-metric, that is, $d_y(y_i,y_j)$ does not satisfy $d_y(y_i,y_j)\leq d_y(y_i,y_k)+d_y(y_k,y_j)$ for $y_i,y_j,y_k$. Thus the inter-point relationships between all the distances in $\bm{\Delta}_{x}$ will not match those in $\bm{\Delta}_{y}$ (for if they did $d_y$ would satisfy the triangle inequality and hence be metric). 

\section{Distance measures}\label{dists}
\subsection{SNPs}

Assume two $P$-dimensional vectors $\bm{x}=(x_{1},\ldots,x_{P})^T$ and $\bm{y}=(y_{1},\ldots,y_{P})^T$ with discrete-valued elements representing minor allele counts at $P$ SNPs. The identity-by-state (IBS) distance measure is defined as
\begin{equation}\nonumber
d_{IBS}(\bm{x},\bm{y})=1-\frac{1}{2P}\sum_{p=1}^Ps(x_{p},y_{p}),
\end{equation}
where $s(x_{p},y_{p})=0$ if $x_{p}=0$ and $y_{p}=2$, or if $x_{p}=2$ and $y_{p}=0$, $s(x_{p},y_{p})=1$ if $x_{p}=1$ and $y_{p}\neq 1$, or if $y_{p}=1$ and $x_{p}\neq 1$, and $s(x_{p},y_{p})=2$ if $x_{p}=y_{p}$. This distance takes values between $0$ and $1$ and is semi-metric.

Genetic distances have also been proposed based on the contingency table containing the frequency that each combination of minor allele counts occurs over the SNPs \citep{selinski2005cluster}; see Table \ref{tab_snp_dist}.
 
\begin{table*}[h!]
\begin{center} 
\caption{Contingency table containing the frequency of a given combination of minor allele count between $\bm{x}$ and $\bm{y}$ over the $P$ SNPs. $m_{kl}$ is the frequency of $\bm{x}$ having $k$ minor alleles and $\bm{y}$ having $l$ minor alleles.}
\label{tab_snp_dist}
\begin{tabular}{c|ccccc}
        \toprule
\backslashbox {$\bm{x}$}{$\bm{y}$}& $0$&$1$&$2$  \\
\midrule
 $0$&$m_{00}$&$m_{01}$&$m_{02}$ \\      
 $1$&$m_{10}$&$m_{11}$&$m_{12}$ \\      
 $2$&$m_{20}$&$m_{21}$&$m_{22}$ \\ 
\bottomrule
\end{tabular}
\end{center}

\end{table*}

The key statistics in this table are the number of complete matches of the minor allele counts, $m^{+}=\sum_{k=0}^2m_{kk}$, and the number of mismatches, $m^{-}=P - m^{+}$, where the total number of possible matches is $P$. Based on these quantities, the following `matching coefficient' distance measures can be defined; the Simple Matching distance 
\begin{equation}
d_{SM}(\bm{x},\bm{y})=1-\frac{m^{+}}{P},\nonumber
\end{equation}
the Sokal and Sneath (SS) distance
\begin{equation}
d_{SS}(\bm{x},\bm{y})=1-\frac{m^{+}}{m^{+}+\frac{1}{2}m^{-}},\nonumber
\end{equation}
and the Rogers and Tanimoto I (RTI) distance 
\begin{equation}
d_{RTI}(\bm{x},\bm{y})=1-\frac{m^{+}}{m^{+}+2m^{-}}.\nonumber
\end{equation}
There is also the Hamman I similarity measure
\begin{equation}
s_{HI}(\bm{x},\bm{y})=\frac{m^{+}-m^{-}}{P},\nonumber
\end{equation}
which can be transformed into a distance measure as follows. Assume $N$ $P$-dimensional minor allele count vectors $\{\bm{x}_i\}_{i=1}^N$; this is required in order to normalize the magnitude of the similarities to the range $[0,1]$. The Hamman I distance between $\bm{x}_i$ and $\bm{x}_j$ is then given by 
\begin{equation}
d_{HI}(\bm{x}_i,\bm{x}_j)=1-\frac{s^*(\bm{x}_i,\bm{x}_j)}{\underset{i,j}{\textrm{max}}\{s^*(\bm{x}_i,\bm{x}_j)\}},\nonumber
\end{equation}
where $s^*(\bm{x}_i,\bm{x}_j)=s_{HI}(\bm{x}_i,\bm{x}_j)+|\underset{i,j}{\textrm{min}}\{s_{HI}(\bm{x}_i,\bm{x}_j)\}|$. This takes values between $0$ and $1$ and is semi-metric.

\subsection{Vectors}
 
Assume two $P$-dimensional real-valued vectors $\bm{x}=(x_{1},\ldots,x_{P})^T$ and $\bm{y}=(y_{1},\ldots,y_{P})^T$. Many measures exist (see, for example, \cite{pekalska2005dissimilarity}), of which a few are provided in Table \ref{vec_dist}, along with their ranges and properties, i.e., whether they are metric or semi-metric. These include the Euclidean, Manhattan, Maximum, Bray-Curtis, Pearson's correlation and the Cosine angle distances.

\begin{table*}[h!]
\begin{center} 
\caption{Commonly encountered distance measures for vector-valued objects. The (M) or (SM) by each distance name indicates whether it is metric or semi-metric.}
\label{vec_dist}
{\bigskip}
\begin{tabular}{l|c|c|c}
        \toprule
Distance&Notation&Definition&Range\\
\midrule
Euclidean (M)&$d_{E}(\bm{x},\bm{y})$&$\sqrt{\left(\bm{x} - \bm{y}\right)^T\left(\bm{x}-\bm{y}\right)}$&$[0,\infty)$\\
&&&\\
Manhattan (SM)&$d_{MAN}(\bm{x},\bm{y})$&$\sum_{p=1}^P\left|x_{p} - y_{p}\right|$&$[0,\infty)$\\
&&&\\
Maximum (SM)&$d_{MAX}(\bm{x},\bm{y})$&$\underset{p}{\textrm{max}}\left\{\left|x_{p} - y_{p}\right|\right\}$&$[0,\infty)$\\
&&&\\
Bray-Curtis (SM)&$d_{BC}(\bm{x},\bm{y})$&$\begin{array}{c}\sum_{p=1}^P\left|x_{p} - y_{p}\right|\\\overline{\sum_{p=1}^P\left(x_{p} + y_{p}\right)}\end{array}$&$[0,\infty)$\\
&&&\\
Mahalanobis (SM)&$d_{MAH}(\bm{x},\bm{y})$&$\begin{array}{c}\sqrt{\left(\bm{x} - \bm{y}\right)^T\bm{S}^{-1}\left(\bm{x}-\bm{y}\right)},\\\textrm{$\bm{S}$ a $P\times P$ covariance matrix, $P<N$} \end{array}$&$[0,\infty)$\\
&&&\\
$\begin{array}{l}\textrm{Pearson's}\\\textrm{correlation (SM)}\end{array}$&$d_{PC}(\bm{x},\bm{y})$&$1-\begin{array}{c}\sum_{p=1}^P\left(x_{p}-\bar{x}\right)\left(y_{p}-\bar{y}\right)\\\overline{\sum_{p=1}^P\left(x_{p}-\bar{x}\right)^2\sum_{p=1}^P\left(y_{p}-\bar{y}\right)^2}\end{array}$,&$[0,2]$\\
&&$\bar{x}=\frac{1}{P}\sum_{p=1}^Px_{p},\quad \bar{y}=\frac{1}{P}\sum_{p=1}^Py_{p}$&\\
$\begin{array}{l}\textrm{Cosine}\\\textrm{angle (SM)}\end{array}$&$d_{PC}(\bm{x},\bm{y})$&$1-\begin{array}{c}\bm{x}^T\bm{y}\\\overline{\left|\left|\bm{x}\right|\right|\left|\left|\bm{y}\right|\right|}\end{array}$,&$[0,2]$\\
&&$\left|\left|\bm{x}\right|\right|=\sqrt{\sum_{p=1}^Px^2_{p}},\quad \left|\left|\bm{y}\right|\right|=\sqrt{\sum_{p=1}^Py^2_{p}}$&\\
\bottomrule
\end{tabular}
\end{center}

\end{table*}

Spearman's correlation distance is defined by applying Pearson's correlation to the ranks of the elements of the vectors, rather than the actual values. In particular, let $\bm{x}_{r}=(x_{r1},\ldots,x_{rP})^T$ and $\bm{y}_{r}=(y_{r1},\ldots,y_{rP})^T$ be the vectors containing the ranks of the elements of $\bm{x}$ and $\bm{y}$, respectively in ascending order (highest value given rank $1$). That is, $x_{rp}$ is the rank of $x_p$, and similarly for $y_{rp}$. If several elements of a given vector are equal, they are assigned a rank equal to the mean of their respective positions in the list of ascending values. For example, for vector $(0.1,0.4,0.4,0.5,-31)^T$, their respective positions are $(4,3,2,1,5)^T$ or $(4,2,3,1,5)^T$, so that the ranks are given by $(4,2.5,2.5,1,5)^T$. Spearman's correlation distance between $\bm{x}$ and $\bm{y}$ is  thus given by
\begin{equation}
d_{SC}(\bm{x},\bm{y})=d_{PC}(\bm{x}_{r},\bm{y}_{r}),\nonumber
\end{equation}
which ranges between $0$ and $2$ and is semi-metric.

The normalized mutual information (NMI) distance is defined as follows.  The $P$ elements of $\bm{x}$ and $\bm{y}$ are considered to be observations of the random variables $x$ and $y$, respectively. Let $p_{x}(\cdot)$ denote the PMF of $\bm{x}$ found by considering the histogram of the elements of $\bm{x}$ with $M$ bins. That is, $p_{x}(i)$ gives the proportion of the elements $\{x_p\}_{p=1}^P$ in the $i^{\textrm{th}}$ bin. We follow \cite{priness2007evaluation} and use the integer value of $\sqrt{P}$ as $M$. Estimate the entropy of $\bm{x}$ by
\begin{equation}
\textrm{E}(x)= -\sum_{m=1}^M p_{x}(m)\textrm{log}\left(p_{x}(m)\right),\nonumber
\end{equation}
and similarly for $\bm{y}$ with PMF $p_{y}(\cdot)$. Estimate the joint entropy of $\bm{x}$ and $\bm{y}$ by considering the joint PMF, denoted  $\{p_{xy}(i,j)\}_{i,j=1}^M$, using a two-dimensional histogram of $\bm{x}$ and $\bm{y}$, as 
\begin{equation}
\textrm{E}(x,y)= -\sum_{m=1}^M\sum_{n=1}^M p_{xy}(m,n)\textrm{log}\left(p_{xy}(m,n)\right).\nonumber
\end{equation}
The NMI distance measure is then given by 
\begin{equation}
d_{NMI}(\bm{x},\bm{y}) =1- \frac{\textrm{E}(x) + \textrm{E}(y)-\textrm{E}(x,y)}{\textrm{max}\left\{\textrm{E}(x),\textrm{E}(y)\right\}},\nonumber
\end{equation}
where the fraction is the NMI \citep{michaels1998cluster}. This distance is bounded by $0$ and $1$ and is semi-metric.

\section{Simulation results}\label{pows}

\subsection{Comparison between GRV and Mantel}

\begin{figure}[h!]
\center
\includegraphics[scale=0.6]{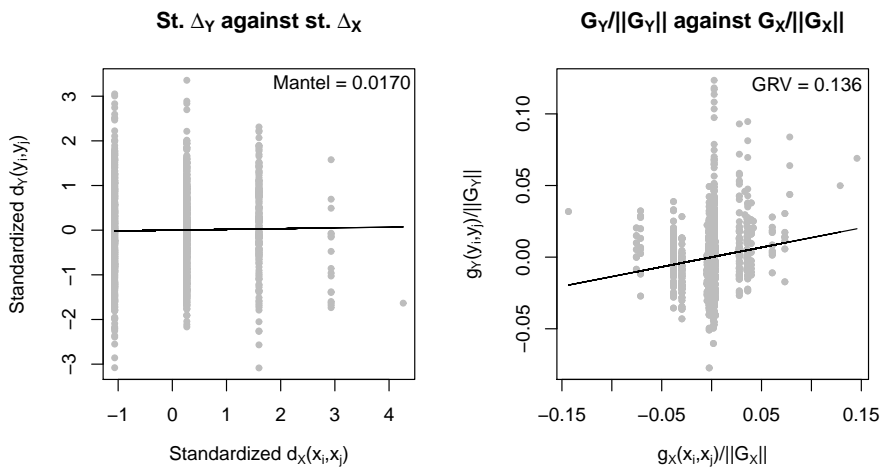}
\caption{Standardized and normalized double-centered distances (gray points) used by the Mantel and GRV statistics, for the IBS and Mahalanobis distances for $N=50$. In each case the gradient of the black line equals the correlation and hence the statistic value. The standardization used by GRV yields a higher correlation than achieved by the standardization used by Mantel.}
\label{cor_plots}
\end{figure}

Since the GRV and Mantel statistics can both be expressed as correlation coefficients, the difference in performance is due to the difference in standardized distance elements used as inputs in each case; a classical standardization is applied to the upper-triangular distance elements by Mantel, and a normalized double-centering is applied to the complete distance matrix by GRV. Figure \ref{cor_plots} demonstrates that, for the data simulated as described in Section $3$ of the article with $N=50$ and with the IBS and Mahalanobis distances, the standardization used by GRV yields a stronger correlation than the standardization used by Mantel. This is an example of how the standardization used by GRV better detects the association inherent in the simulated data.

\begin{table}[h!]
\begin{center} 
\caption{Mean (and standard deviation) of the proportion of null p-values less than or equal to the given significance levels for $N=\{30,50,70\}$.}
\label{nom_sigs}
\begin{tabular}{llllllllll}
        \toprule
&& \multicolumn{8}{c}{Significance level ($\%$)}\\
N&& \multicolumn{3}{l}{$1$}&\multicolumn{3}{l}{$5$}&\multicolumn{2}{l}{$10$}\\
\midrule
$30$&&$0.010$&$(0.007)$&&$0.051$&$(0.014)$&&$0.100$&$(0.020)$\\
$50$&&$0.010$&$(0.006)$&&$0.050$&$(0.016)$&&$0.101$&$(0.020)$\\
$70$&&$0.011$&$(0.006)$&&$0.051$&$(0.017)$&&$0.102$&$(0.021)$\\
\bottomrule
\end{tabular}
\end{center}
\end{table}

\begin{figure}[h!]
\center
\includegraphics[width=1.0\textwidth]{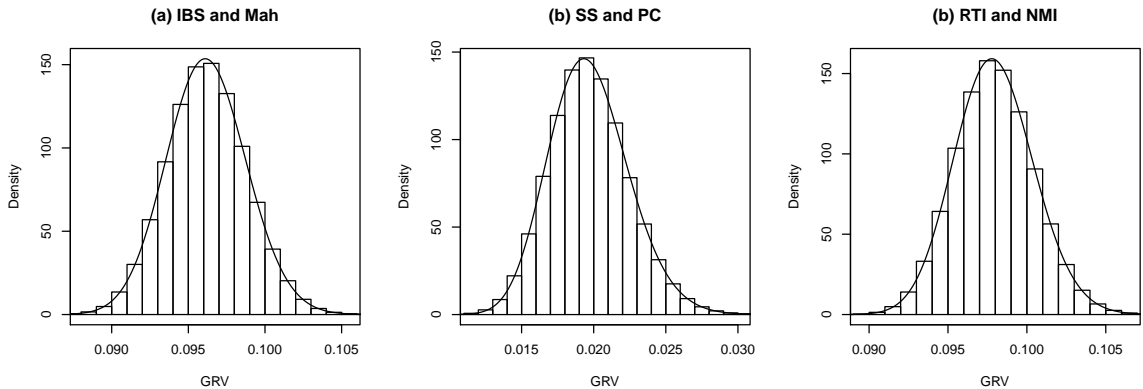}
\caption{Sampling distributions of the GRV statistic obtained using $10^6$ Monte Carlo permutations and the proposed approximate PDF applied to a pathway of the real data including a SNP-set comprised of $2092$ SNPs and a transcript-set comprised of $51$ transcripts. (a) The IBS distance is applied to the SNP data and the Mahalanobis (Mah) distance is applied to the gene expression data. (b) The SS distance is applied to the SNP data and the Pearson's correlation distance (PC) distance is applied to the gene expression data. (b) The RTI distance is applied to the SNP data and the normalized mutual information (NMI) distance is applied to the gene expression data.}
\label{dist_plots}
\end{figure}

\subsection{Attainment of nominal significance level of GRV}

We demonstrate that the GRV test attains the nominal significance level of a given test for a range of significance levels by using a Monte Carlo procedure. We generate paired data inspired by the eQTL application setting used in the power study.

For $100$ Monte Carlo runs, $200$ paired datasets are generated as described in Section $3$ of the article for $P=2$, $Q=10$, and $N=\{30,50,70\}$, except with $\bm{y}_i=\bm{e}_i$ so that $\bm{y}_i$ is not dependent on $\bm{x}_i$, and the IBS and Mahalanobis distance measures applied. Over all runs the mean and standard deviation of the proportion of p-values less than or equal to significance levels of $1\%,5\%,10\%$ are monitored. These are presented in Table \ref{nom_sigs}, where it is clear that the nominal significance levels are attained.

%

\section{Illustration of the null distribution of GRV for real data}\label{data}

In order to illustrate how the null distribution compares with the permutation distribution when applied to real data, we consider a subset of the cancer data. For a SNP-set comprised of $2092$ SNPs and a transcript-set comprised of $51$ transcripts, each of the following combinations of SNP and gene expression distances were applied: (a) IBS and Mahalanobis, (b) SS and Pearson's correlation, and (c) RTI and normalized mutual information. For each we obtained the approximate null distribution of the GRV statistic and the permutation distribution using $10^6$ Monte Carlo permutations. These are given in Figure \ref{dist_plots}, where we see that the approximate distribution appears to provide a good fit to the permutation distributions.


\bibliography{grv_bioinf_bib_final}
\bibliographystyle{natbib}

\end{document}